\documentclass[12pt]{article}
\usepackage{epsfig}

\def\beq{\begin{equation}}
\def\eeq#1{\label{#1}\end{equation}}
\def\eeqn{\end{equation}}

\def\beqa{\begin{eqnarray}}
\def\eeqa#1{\label{#1}\end{eqnarray}}
\def\eeqan{\end{eqnarray}}

\let\bar=\overbar

\def\Dslash{\not{\hbox{\kern-4pt $D$}}}
\def\dslash{\not{\hbox{\kern-2pt $\del$}}}

\def\msb{{\bar{\ssstyle M \kern -1pt S}}}

\def\Title#1{\begin{center} {\Large {\bf #1} } \end{center}}

\begin{document}

\Title{Godot and the New Physics}

\bigskip\bigskip


\begin{raggedright}  

{\it Guy Wilkinson\index{Wilkinson, G.}\\
University of Oxford, \\
Denys Wilkinson Building,\\
Keble Road,\\
Oxford, OX1 3RH, \\
UNITED KINGDOM}
\bigskip\bigskip
\end{raggedright}







\section{Waiting for Godot: a Parable for the Flavour Physics Community?}

Samuel Beckett's play of 1949 is a landmark of modern theatre.
Two tramps, Vladimir and Estragon, await the arrival of the mysterious
Godot.  He does not come, although other sinister characters pass through.  
The tramps pass their time in meaningless
activities and talk. Described by Beckett as a `tragicomedy in two acts', the play
is conventionally regarded as a tale of existentialist angst focused on the futility 
of the human condition.  Other readings are of course possible.  The interpretation
we suggest here is that the play is a parable of the search for New Physics in the
flavour sector.  
\begin{itemize}
\item{{\bf Why are we here?  Godot as the New Physics} \\
Like Vladimir and Estragon, we hope we know why we are here.  As Vladimir says ``But that is not
the question. Why are we here, {\it that} is the question.  And we are blessed in this, that we happen
to know the answer. Yes in this immense confusion one thing alone is clear.  We are waiting for 
Godot to come.''~\cite{GODOT}  We know why we are doing what we are doing. We are awaiting the arrival of the
New Physics, which we are convinced must manifest itself in flavour observables.  This conviction
can sometimes make us a little too enthusiastic.}
\item{{\bf ``It's Godot, we're saved!''} \\
The tension of waiting sometimes overwhelms our heroes -- ``That's to say... you understand... 
the dusk... the strain... waiting... I confess... I imagined... for a second...'' Estragon apologises,
after having misidentified another character for Godot.  The flavour community has made similar
bold statements.  For example, in the early era of the B-factories the data supported a different
value for the value of $\sin 2 \beta$ as measured in processes involving box and $b\to s$ Penguin
diagrams, such as $B^0 \to \phi K^0_S$, compared with those involving box and tree diagrams,
such as $B^0 \to J/\psi K^0_S$ (see Fig.~\ref{fig:phiks}). This discrepancy provoked much excitement,
excitement which has now largely dissipated as the two sets of results have become much more consistent. }
\begin{figure}[h]
\begin{center}
\vspace{1.8cm}
\epsfig{file=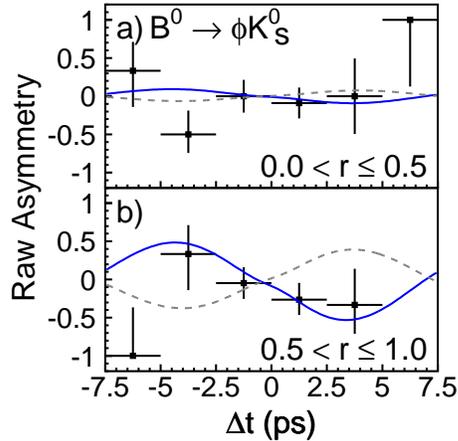,width=0.4\textwidth}
\caption{Belle results for the CP-asymmetry in $B^0\to \phi K^0_S$ as measured with 140~${\rm fb^{-1}}$
of data~\cite{BELLEOLDPHIKS}.  Plot a) represents events with low tagging purity, plot b) events
with high tagging purity. The blue solid curve is the fit result, the dotted black curve the
Standard Model expectation. There is a $3.5 \sigma$ discrepancy between the two. With
more data analysed from both BABAR and Belle, the discrepancy is now only $1.3 \sigma$~\cite{HFAG}.}\label{fig:phiks}
\end{center}
\end{figure}
\item{{\bf ``Has he a beard, Mr Godot?''} \\
When a boy, supposedly familiar with Godot, answers this question from Vladimir in the affirmative, the tramp
then asks its colour: ``Fair or... ...or black?''. Boy: ``I think it's white, Sir''. {\it Silence}. Vladimir: ``Christ have mercy 
on us!''.
This interchange highlights another challenge that we face. As we search for the New Physics we do not know quite what
we are looking for -- SUSY, Little Higgs or maybe something else?}
\item{{\bf But we persevere} \\
Flavour physicists are by nature stubborn people.  The absence (so far) of a clear signpost to a higher
theory from flavour observables at existing and past facilities does not dissuade us from planning for 
still more sensitive measurements. We know out quest is well motivated.
Vladimir: ``What are you insinuating? That we've come to the wrong 
place?'', Estragon: ``He should be here.'', Vladimir: ``He didn't say for sure he'd come'', 
Estragon: ``And if he doesn't come?''. Vladimir: ``We'll come back tomorrow'',  Estragon: ``And then
the day after tomorrow'', Vladimir: ``Possibly.''}
\end{itemize}

Note that Vladimir is more interested in his prospects of meeting Godot `tomorrow', rather than 
`the day after tomorrow', and the same spirit will guide this review.  In the world of experimental
particle physics `tomorrow' may be regarded as the coming five years, and the `day after tomorrow'
the time beyond.  In this survey, therefore, we will focus on those areas where we have may have genuine reason
to hope where New Physics may appear in the coming five years, or at least where advances are 
expected which are necessary for such an event to occur.  The choice of topics will be subjective,
but will be informed by the many excellent presentations given at this conference.   By excluding
the `day after tomorrow' we are choosing not to discuss the exciting prospects at those experiments 
still on the planning board, such as the {\it upgraded} LHCb experiment, or a very high luminosity  
$e^+e^-$ flavour-factory. There will be time enough to explore these at future conferences.

\section{Time Dependent CP-violation}

Mixing was first established in the neutral $B$ system in 1987~\cite{MIXDISC}. It took another
six years before the actual oscillations themselves were resolved for $B^0$ mesons~\cite{ALEPHMIX}.
Impressive though this feat was, it was merely a staging post on the journey towards the 
real prize of observing mixing-induced CP-violation, a goal which was attained by the $B$-factory
experiments in 2001~\cite{CPVINBD}.  

\begin{figure}
\begin{center}
\epsfig{file=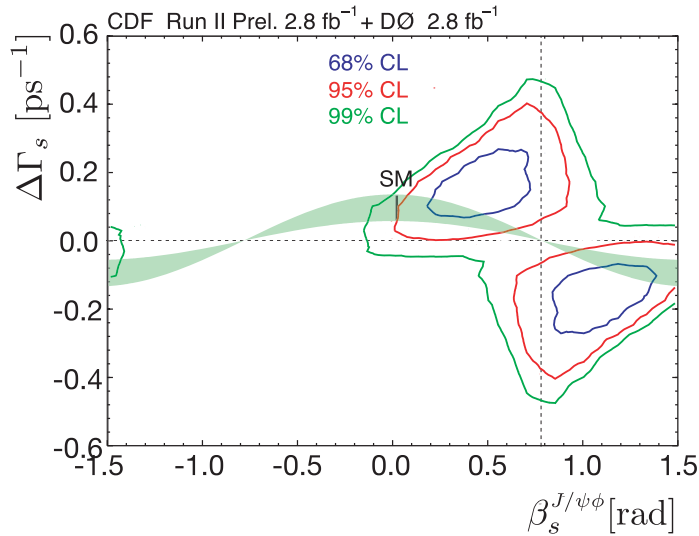,width=0.6\textwidth}
\caption{CDF and D0 combined profile likelihood as confidence contours of the 
CP-violating phase $\beta_s^{J/\psi \phi}$  
in $B^0_s \to J/\psi\phi$ (referred to as $\phi_s$ in the main text)
and the width splitting between the mass eigenstates, 
$\Delta \Gamma_s$. The Standard Model
expectation and uncertainty is indicated by the black line.  The sinusoidal green band indicates that region
allowed in New Physics models.~\cite{EPSPHIS}}\label{fig:tevphis}
\end{center}
\vspace*{-0.5cm}
\end{figure}

Is a similar story now unfolding with $B^0_s$ mesons?  
The very rapid oscillations were only resolved in 2006~\cite{BSOSC}, but already CDF and D0
are now searching for CP-violation in the gold-plated mode $B_s \to J/\psi \phi$.
In the Standard Model the CP-violating phase $\phi_s$~\footnote{Note that sometimes in the literature the symbol
$\phi_s$ is used instead for the different phase which is accessed through measuring the flavour-specific asymmetry in
$B^0_s$ decays.} characterising this effect
is both precisely predicted and known to be very small 
($\phi_s^{SM} = -0.0368 \pm 0.0017$ radians~\cite{CKMFITTER}).  The early analyses~\cite{TEVCPVEARLY},
however, although still rather insensitive, hint at a larger value.  This hint has been seized on
by certain commentators~\cite{UTFITPHIS} as heralding the coming of the New Physics,  but
such an interpretation relies on the reliable averaging of the results from the two experiments,
which in the case of the confidence level contours in $\Delta \Gamma_s - \phi_s$ space,
where $\Delta \Gamma_s$ is the lifetime splitting between the mass eigenstates, is a very non-trivial exercise.
The most recent combination~\cite{EPSPHIS} performed by the the Tevatron collaborations themselves,
and displayed in Fig.~\ref{fig:tevphis},
indicates that the  preferred central value is $2.1 \sigma$ 
away from the Standard Model prediction.  This is intriguing, but for sure Godot has not yet arrived.

\begin{figure}
\vspace{1.5cm}
\begin{center}
\epsfig{file=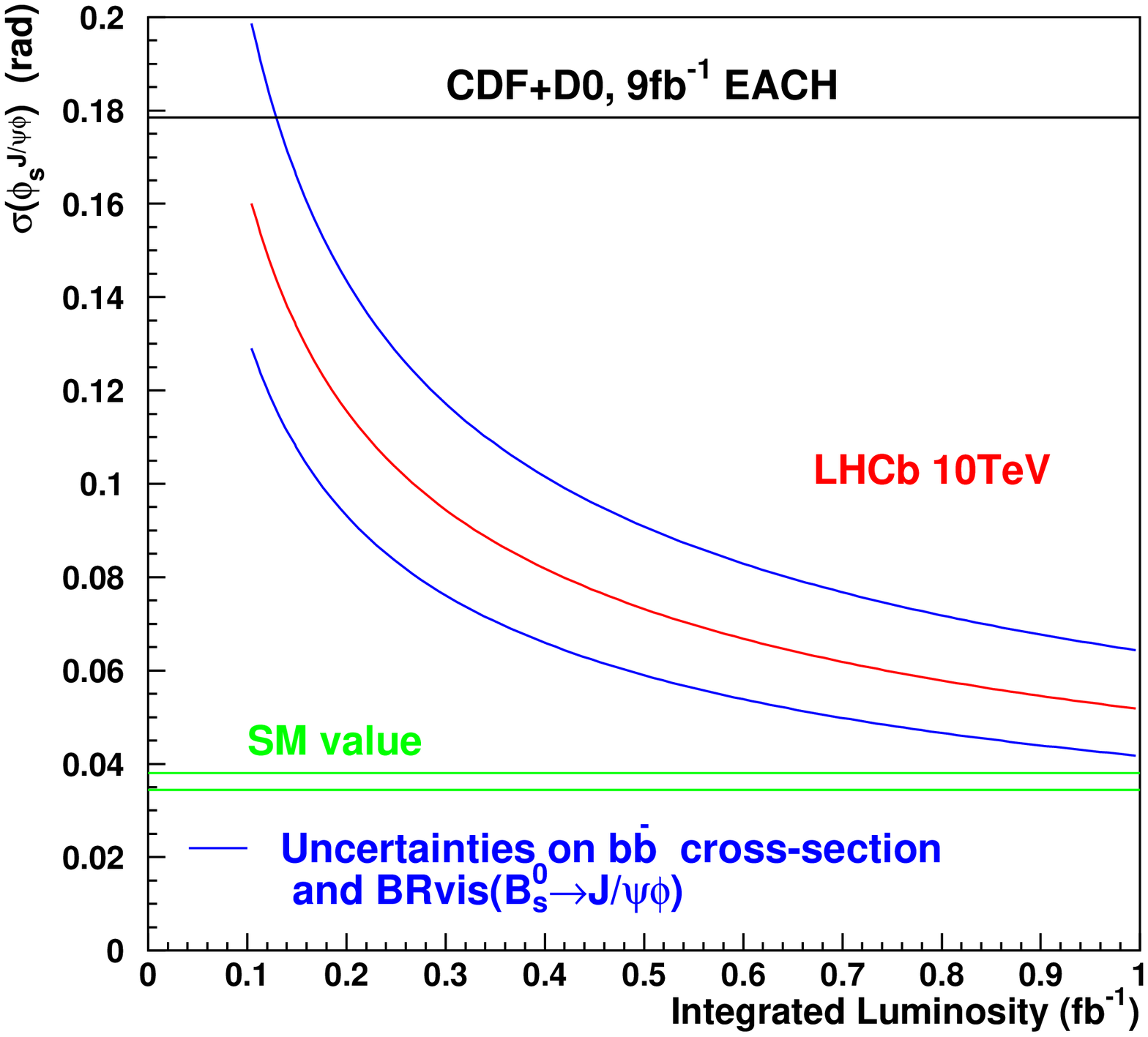,width=0.48\textwidth}
\epsfig{file=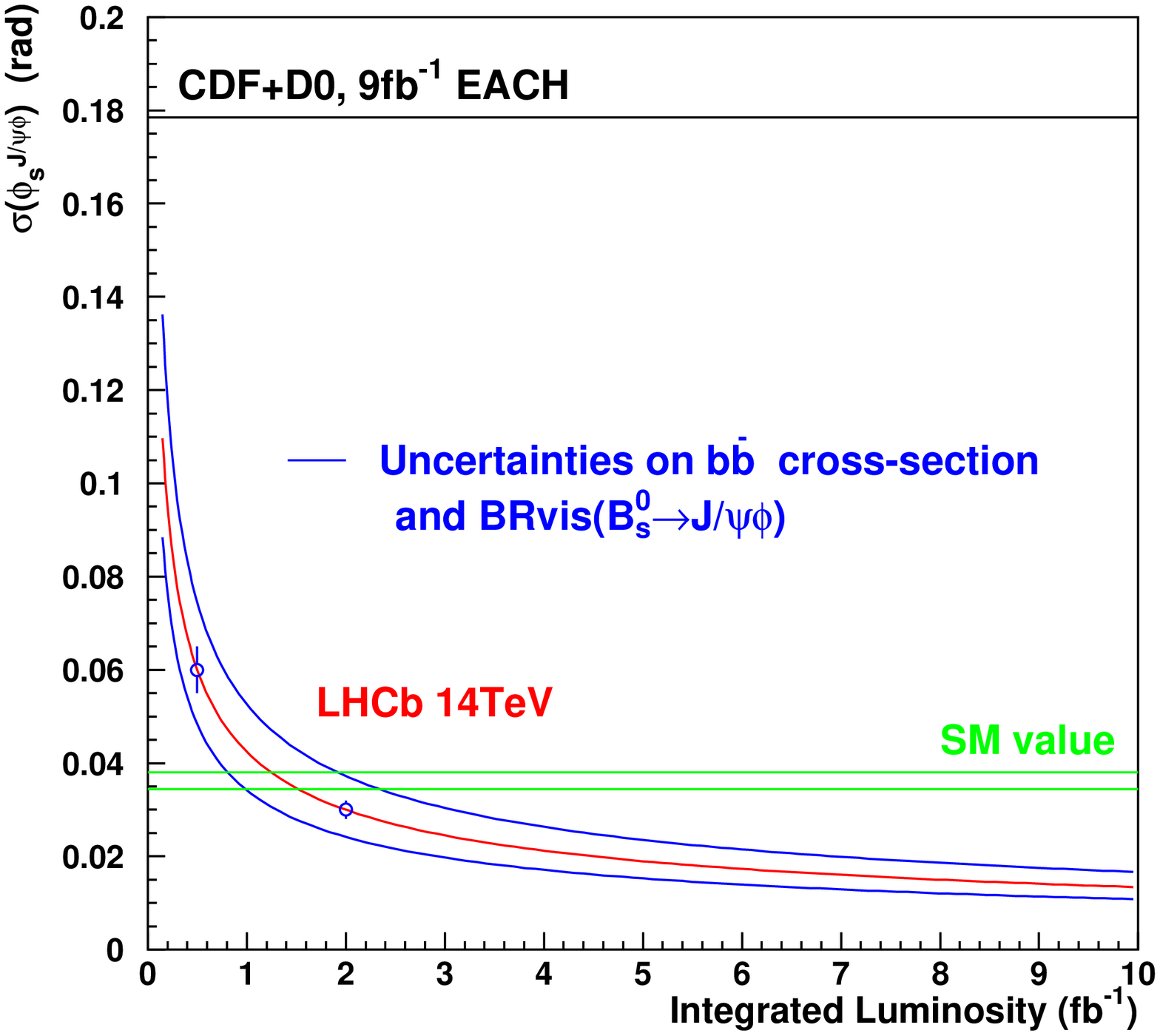,width=0.48\textwidth}
\caption{Expected LHCb sensitivity to CP-violating phase $\phi_s^{J/\psi \phi}$  
in $B^0_s \to J/\psi\phi$ (referred to as $\phi_s$ in the main text) as 
a function of integrated luminosity.  Left: prospects in 2010 at $\rm E_{cm} = 10\, TeV$. Right: prospects at nominal LHC energy.
The Tevatron estimates are naive scalings from the present results.}\label{fig:lhcbphis}
\end{center}
\end{figure}

Nonetheless, it is exciting to appreciate that if the present results for $\phi_s$ are indeed
centred on the true value, then discovery of the New Physics is rather imminent.  The present Tevatron
results are based on 2.8$~{\rm fb^{-1}}$ of data per experiment and by the end of 2010
three to four times this data size will be available, which would give CDF and D0
combined a very good chance of a five-sigma observation.  Next year also, the first results
are expected from the LHC.  As is shown from Fig.~\ref{fig:lhcbphis}, LHCb is expected to match the anticipated
Tevatron performance with around 200~${\rm pb^{-1}}$ of data, which is a realistic integrated 
luminosity goal for 2010.  
  
It may be, of course, that improved measurements of $\phi_s$ will exclude
the initial indications of a very large value. Such an outcome
would not diminish the importance of $\phi_s$ as an observable which
a priori has excellent potential to reveal the contributions of New Physics.  
To this end, it will be essential
to push the sensitivity of the analyses down to the level of the
Standard Model expectation, and beyond.   
Using  $B^0_s \to J/\psi \phi$ events alone, LHCb has the capability
to attain this precision with less than 2~${\rm fb^{-1}}$ of data, 
a sample which should be accumulated
within the first two or three years of LHC operation.  Other channels
may allow for the experimental uncertainty to be reduced still further.
A promising candidate is 
$B^0_s \to J/\psi f_0(980)$ with $f_0(980)\to \pi^+\pi^-$.
As the final state is a CP-eigenstate, in contrast to the vector-vector nature of $J/\psi \phi$, no
angular analysis is necessary, and so the intrinsic sensitivity per event is higher.
The relevant question is then what the relative branching ratio of this decay is
with respect to $B_s \to J/\psi \phi$, $\phi \to K^+K^-$. 
In~\cite{SHELDONPSIF0} it is suggested that this ratio should be the same
as that of the decays $D_s^+ \to f_0 e^+ \nu_e$, $f_0 \to \pi^+\pi^-$ and
$D_s^+ \to \phi e^+ \nu_e$, $\phi \to K^+K^-$ when extrapolated to
$q^2=0$.  CLEO-c has made the first measurement 
of this ratio and found it to be $42 \pm 11\%$~\cite{CLEOCF0}, which bodes 
well for the contribution of the $B_s \to J/\psi f_0(980)$ in the $\phi_s$ determination.
It is hoped that the Tevatron will soon make a direct measurement of this branching fraction.

It should be remembered that there is another opportunity for mixing measurements
to reveal CP-violating contributions from beyond the Standard Model.
One of the surprises of the last two years has been the discovery of 
oscillations in the $D^0-\bar{D^0}$ system.  This observation has
already been used to constrain or exclude many models of New Physics~\cite{GOLOWICH}.
The charm mixing parameters, $x$ and $y$, are now both known to be $\sim 1\%$, with a
relative precision of around $25\%$~\cite{HFAG}. Although small,
the oscillation is at the higher end of the expected spectrum and gives hope
for observing any mixing related CP-violation effects, which are 
pre-scaled by factors of either $x$ or $y$~\cite{DMIXPHENOM}.  In the 
Standard Model these effects are utterly negligible,  but significant
contributions may occur in many New Physics models (see for example~\cite{NEUBERT, BIGICPV}).
Sensitivity to such effects can be attained by merely dropping the
assumption of CP-conservation in the mixing studies.  Current measurements
yield the results $|q/p|_D = 0.86^{+0.17}_{-0.15}$ 
and $\phi_D = (-8.8^{+7.6}_{-7.2})^\circ$~\cite{HFAG},
where the CP-conserving predictions are $1$ and $0^\circ$ respectively. 
Significant improvements are expected in the coming few years from the $B$-factories,
the Tevatron and LHCb.   The same facilities will also search for direct CP-violation,
particularly in singly Cabibbo-suppressed decays, which represent another promising
area for beyond-the-Standard-Model physics to manifest itself~\cite{SCSD}.

\section{The Unitarity Triangle: measuring $\gamma/\phi_3$}

The least well known angle of the unitarity triangle is $\gamma$ ($\phi_3$).
Even the degree of imprecision in our knowledge of $\gamma$ is itself not well established:
the UTfit collaboration has performed a global combination of existing measurements
and obtains a value of $(78\pm 12)^\circ$~\cite{UTFIT}, whereas CKMFitter
calculates the world average to be $(70^{+27}_{-30})^\circ$~\cite{CKMFITTER}.
The desire to make more a precise measurement of this angle is further
motivated by the fact that it is the only CP-violating parameter that can
be measured through tree-level processes, and therefore provides a Standard
Model benchmark, largely immune to New Physics effects, against which other
observables can be compared.  Thus a much improved determination of $\gamma$
has highest priority in unitarity triangle studies over the coming 5 years or so.

The most powerful way to determine $\gamma$ directly, and the one which is exploited in
all existing publications from the $B$-factories, is the so-called `$B\to DK$' family of measurements.
Interference between $B^- \to D^0 K^-$ and $B^- \to \bar{D^0}K^-$ occurs if the
$D^0$ and $\bar{D^0}$ are reconstructed in a common final state.  Such interference
picks out the relative phase difference, $\delta_B - \gamma$, between the two processes,
where $\delta_B$ is a CP-invariant strong phase.  
Comparison between $B^-$ and $B^+$
decays will exhibit differences in  the event rates or kinematical distributions (for 
example Dalitz plots in the case of $\ge$3 body D-decays), which enable $\gamma$ to be
determined.  There are many $D$ decay channels which can be harnessed for this purpose.
The most promising modes include CP-eigenstates (eg. $K^+K^-$), Cabibbo favoured/doubly 
suppressed decays (eg. $K^\pm\pi^\mp$) and self-conjugate states (eg. $K_S \pi^+\pi^-$).
Apart from in the CP-eigenstate case, the CP-violating observables will have a 
dependence not only on $\gamma$, $\delta_B$ and $r_B$ (the relative size of the
interfering diagrams), but also on strong
phase differences associated with the $D^0/\bar{D^0}$ decay.  These differences
must also be extracted in the analysis, or constrained from external sources.
By including as many channels as possible in the analysis improved
sensitivity is obtained on the common unknowns $\delta_B$, $r_B$ and $\gamma$ itself.

The prospects for an improved determination of $\gamma$ are good.  The $B$-factories
can both update existing analyses with their full $\Upsilon(4S)$ datasets,
and add new channels.   The Tevatron has shown its potential
for contributing to this programme~\cite{CDFGAMMA}.  The $\gamma$ determination
is a principal goal of LHCb, where methods such as time-dependent $B_s \to D_s^\pm K^\mp$
studies can be used to augment the $B\to DK$ strategies.  It is estimated that
LHCb can reach a precision of $2-3^\circ$ with 10~${\rm fb^{-1}}$ of data~\cite{AKIBA}.

The $\gamma$ measurement benefits greatly from the synergy that exists
between facilities.
Information on the strong-phase differences 
associated with the $D^0/\bar{D^0}$ decays which are needed in the
$B \to DK$ analyses can be obtained in a model independent way from
studying the behaviour of quantum-correlated D-mesons produced at the $\psi(3770)$.
Such events have been accumulated at CLEO-c and used to measure the strong
phase differences and related quantities in two, three and four body D-decays~\cite{CLEOCGAMMA}.
Very soon larger samples of $\psi(3770)$ data will become available at BES-III~\cite{BRIERE} to repeat
and extend these analyses.

\section{Rare Decays: looking for Godot in CP-conserving processes}

Study of the branching ratio and kinematical properties of rare heavy flavour decays
have long played an important role in New Physics searches.  For example the branching 
ratio of $b \to s \gamma$ imposes severe constraints in SUSY-parameter space.
The significance of such studies will not diminish over the coming few years.  Here we focus on two of the
most promising candidate channels.

The decay $B \to K^{(*)}\ell^+\ell^-$, which proceeds through a $b\to s$ loop transition, 
is a system which provides a host of powerful observables
which are sensitive to non-Standard Model contributions~\cite{KSTARLL1,KSTARLL2},
in particular the helicity structure of any New Physics couplings.  
One of the most interesting of these observables available in $B^0 \to K^{*0}\ell^+\ell^-$
decays is the forward-backward asymmetry  of the angle between the lepton and the B-meson in the di-lepton rest frame.
This asymmetry is expected to evolve with, $q^2$, the invariant mass of the lepton pair in
a manner which differs between the Standard Model and many New Physics
scenarios.  In particular, in the Standard Model there exists a `zero-crossing point' where the 
asymmetry changes sign, the position of which has rather little theoretical uncertainty 
($q^2 = 4.36^{+0.33}_{-0.31}\, {\rm GeV}^2$~\cite{BENEKE}).
Locating the position of this asymmetry is a key goal in flavour physics.

\begin{figure}[h]
\begin{center}
\epsfig{file=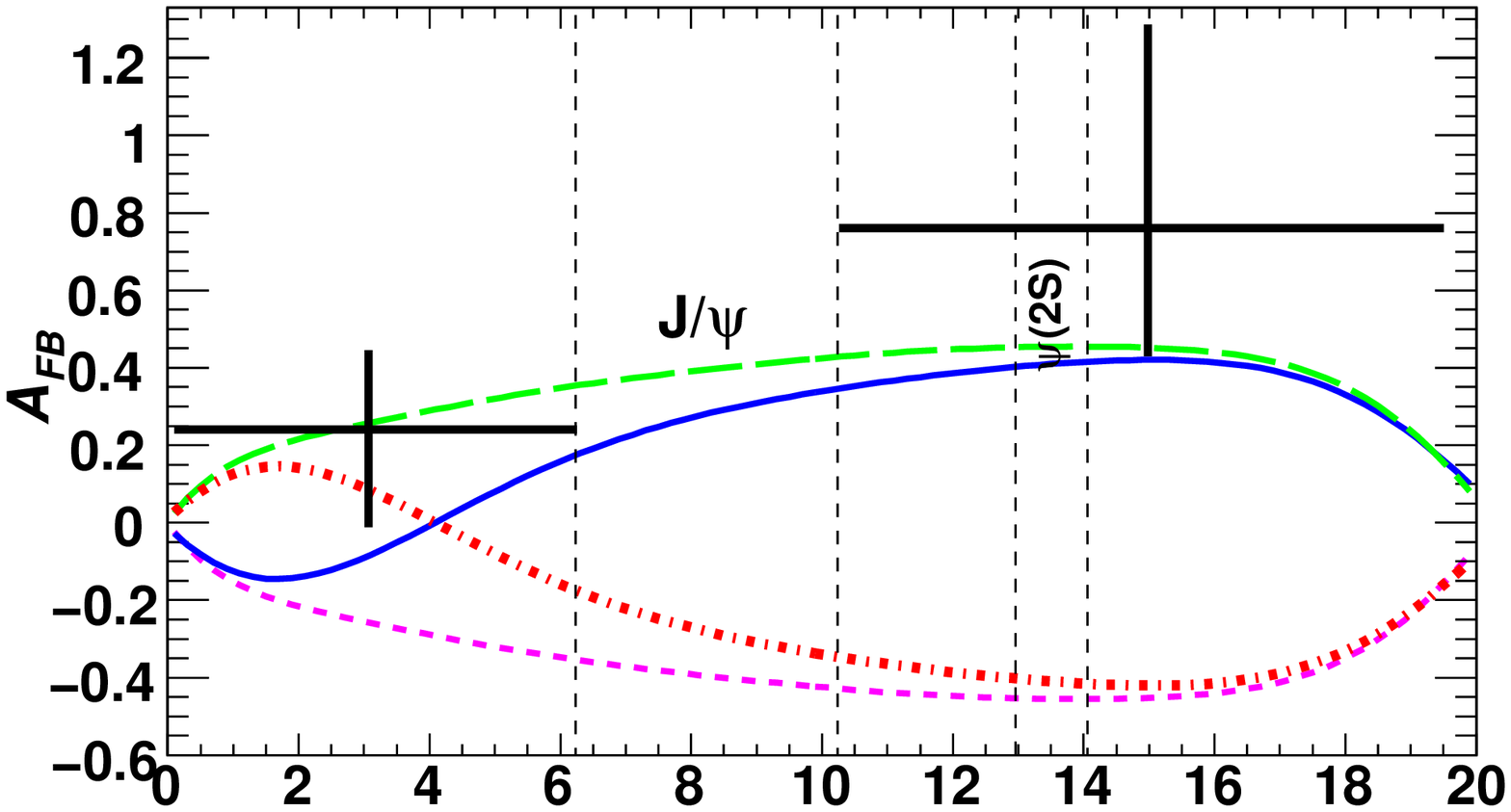,width=0.48\textwidth}
\epsfig{file=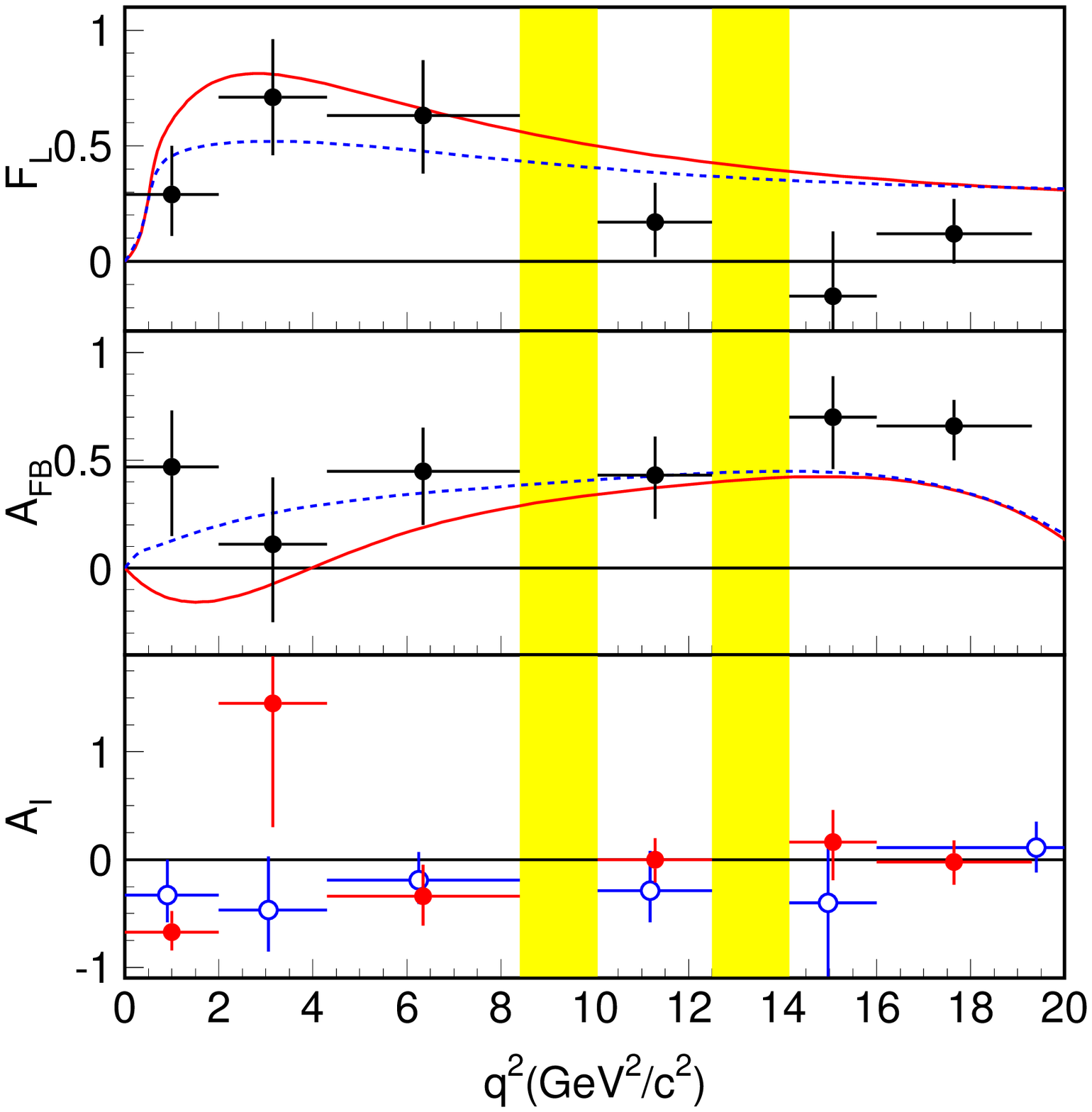,width=0.48\textwidth}
\caption{B-factory results for $B \to K \ell^+\ell^-$ as a function of the invariant mass squared of the
dilepton system.  In all plots the charmonium resonance regions have been excluded.  Left:
BABAR $A_{FB}$ in  $B^0 \to K^{*0} \ell^+\ell^-$; the blue solid curve is the Standard Model prediction, 
the other curves are the expectations when the signs of certain effective Wilson coefficients are swapped 
(eg. long-dashed green is $C_7^{eff}=-C_7^{eff}$)\cite{BABARKSTARLL2}. Right middle: Belle $A_{FB}$ results, with red solid curve
the Standard Model prediction, and the blue dotted curve the expectation with $C_7^{eff}=-C_7^{eff}$\cite{BELLEKSTARLL}.
(Right top: the $K^*$ longitudinal polarisation fraction, with the curves having the same meaning as for $A_{FB}$ plot.
Right bottom: Belle isospin asymmetry for $K^*\ell^+\ell^-$ (closed circles) and $K \ell^+\ell^-$ (open circles).)}
\label{fig:kstar_afb}
\end{center}
\end{figure}

\begin{figure}[h]
\begin{center}
\epsfig{file=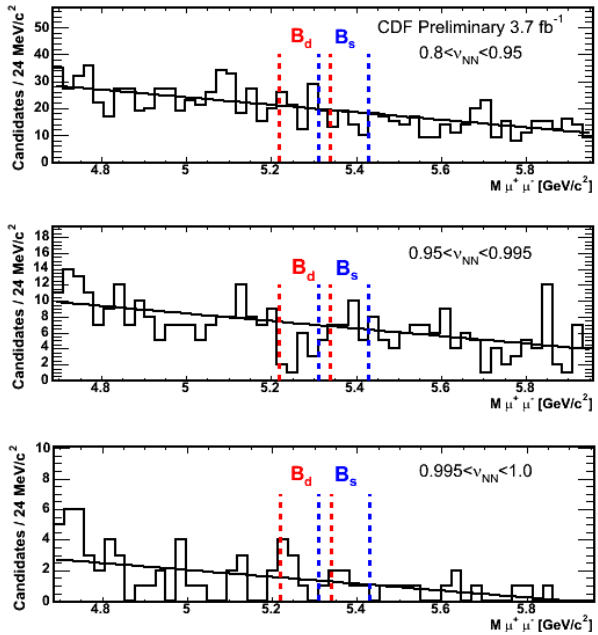,width=0.6\textwidth}
\caption{CDF $B^0_{s} (B^0) \to \mu^+\mu^-$ search with 3.7~${\rm {fb^{-1}}}$~\cite{CDFBSMM}.  Invariant mass of $\mu^+\mu^-$ for 
different intervals of the neural net output used to characterise the event.  The lowest 
plot ($\nu_{NN} > 0.995$) is most signal like.  The search windows are indicated for
both $B^0 \to \mu^+\mu^-$ and for $B^0_s \to \mu^+\mu^-$.}\label{fig:cdf_mumu}
\end{center}
\end{figure}

The $B$-factory experiments have analysed the majority of their collected data.
This has enabled them to present results based on a few hundred signal events~\cite{BABARKSTARLL1,BABARKSTARLL2,BELLEKSTARLL}.
These statistics are inadequate for any conclusions yet to be drawn, but it is 
interesting to note that no indication of a crossing point is yet evident (see Fig.~\ref{fig:kstar_afb}).  
In order to learn more it will be necessary to wait for results from LHCb.  With a dataset
of $\sim$~200~${\rm pb^{-1}}$, perhaps achievable in the first year of operation,
it will be possible to approximately double the existing world-sample of events.
Firm conclusions, however, will only be possible with the luminosities foreseen
from 2011 onwards.  With 10~${\rm fb^{-1}}$, around 5 years of data, 
it will be possible to determine
the zero-crossing point of the asymmetry to the present level of theoretical uncertainty.
To return to our original theme: in $B \to K^{(*)}\ell^+\ell^-$ Godot may well come tomorrow,  
but it is unlikely to be in the morning.
The impatient are advised to focus their attention of the channel $B^0_s \to \mu^+\mu^-$.
This is the $B$-physics rare decay {\it par excellence}.  In the Standard Model the
branching ratio is both highly suppressed and precisely predicted
($\mathcal{B}(B^0_s \to \mu^+\mu^-)=(3.35 \pm 0.32) \times 10^{-9}$~\cite{BLANKE}),
while in many New Physics models of interest substantial
enhancements occur.  For example, calculations made in the context of MSSM point 
to $\mathcal{B}(B^0_s \to \mu^+\mu^-)\sim 2 \times 10^{-8}$ as being
a likely value~\cite{WMAP}.   This is only a little below the best present limit from
CDF of $4.3 \times 10^{-8}$ (95\% C.L.)~\cite{CDFBSMM}, achieved with 3.7~$\rm fb^{-1}$
of data (see Fig.~\ref{fig:cdf_mumu}). With the integrated luminosity which will be available to both experiments within
the coming year it is clear that at the least very powerful constraints can be placed in
SUSY parameter space.   Furthermore, LHCb will, in principle, be
able rapidly to attain a similar sensitivity, reaching the $2 \times 10^{-8}$ level
with the data sample expected in 2010.  Thus if nature conforms to the MSSM
the observation of an enhanced rate of $B_s \to \mu^+\mu^-$ events in the next
one or two years may very well be the first indication of this fact.

\section{Towards CP-violation in the lepton sector}

The choice of topics in this review ($B$ and $D$ physics)
has largely been  motivated by the imminent (re-)start of the 
LHC, and the harvest of results now coming from the Tevatron.
It must be remembered, however, that equally important and complementary
flavour physics studies are underway in other areas. 
These include the search for very rare kaon decays,
LFV muon and tau decays and the drive to improve
the experimental limits on nuclear and lepton EDMs.  
We pass over these topics through lack of time.  Instead, we make brief mention
of that subject which is receiving most effort and attention in the 
domain of neutrino physics, and the one where significant progress is
expected in the coming half-decade.

A host of experiments are approved (or indeed entering data-taking) with
the goal of better constraining, and hopefully measuring,  the 
PMNS mixing angles $\theta_{13}$.  This angle is known to be small ($\sin^2 \theta_{13} \le 0.032$~\cite{PDG08})
and is of particular interest as it controls the magnitude of any CP-violating
observables associated with $\delta$, the phase of the PMNS matrix. Observing
CP-violation in the neutrino sector and measuring $\delta$ are long-term
goals in neutrino physics which are the focus of next-generation experiments,
and therefore are, in Godot-parlance, unlikely to be reached until the `day after tomorrow'. 

The $\theta_{13}$ programme will be conducted in parallel by reactor $\nu_e$ disappearance
experiments (Double Chooz, Daya Bay and RENO) and off-axis $\nu_e$ appearance 
in $\nu_\mu$ superbeam experiments (T2K and NOVA).  Each project has a typical
sensitivity to $\sin^2 2\theta_{13}$ in the range $0.001-0.01$, and the two classes
of technique have very different systematics.  For example the signal is expected
to be larger in superbeam experiments, but its interpretation is complicated
by the presence of matter effects, which are not present in the reactor 
approach.

\begin{figure}
\vspace*{-2.0cm}
\begin{center}
\epsfig{file=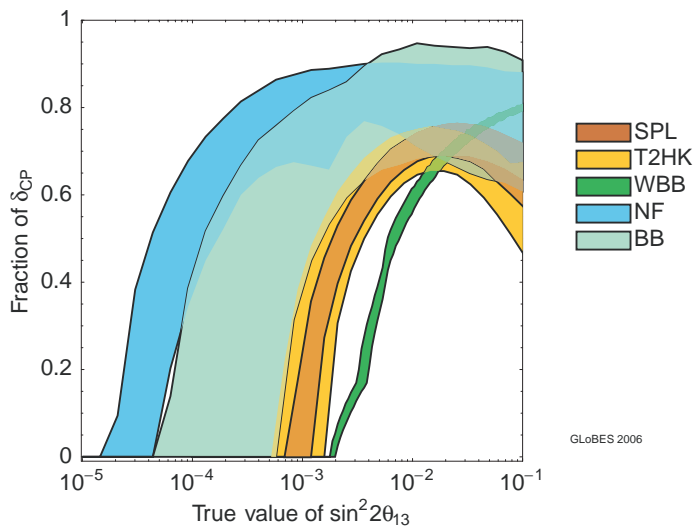,width=1.1\textwidth}
\caption{Discovery reach to CP-violation of various proposed future neutrino facilities 
as a function of the fraction of all possible values of the PMNS CP-violating phase $\delta_{CP}$ and the
true value of $\sin^2 2\theta_{13}$~\cite{ISS}.  In the area to the right of the
bands CP-violation can be established at the $3 \sigma$ confidence level. The 
right most edge of each band represents 
the performance of a conservative design, and the left most an optimised design 
for the facility in question. 'NF' signifies neutrino factory; 'BB' signifies beta-beam.
The other 3 options are superbeam experiments: `SPL' signifies the CERN 4~MW Super Proton Linac
upgrade; `T2HK' signifies the 4 MW, 50~GeV JPARC upgrade; `WBB' signifies the wide band,
1~MW, 28~GeV AGS project.}\label{fig:nu_cpv}
\end{center}
\end{figure}

Improved knowledge of the magnitude of $\theta_{13}$ is necessary for planning
the next generation neutrino experiments, as is indicated in Fig.~\ref{fig:nu_cpv}. 
For example, if it were known that $\sin^2 2\theta_{13}$ is $\sim 10^{-2}$ then so-called 
`wide-band' superbeams would have good sensitivity to CP-violating effects,
while at lower values the more ambitious `beta-beam' or neutrino-factory projects
would be necessary~\cite{ISS}. It is hoped that rather soon we will be in 
a position to decide on which approach should be adopted.

\section{Conclusions}

For the protagonists of the play a happy outcome is contingent on the arrival of Godot.  
Vladimir: ``We'll hang ourselves tomorrow. ({\it Pause}) Unless
Godot comes.'' Estragon: ``And if he comes?''  Vladimir: ``We'll be saved.''  
Flavour physicists are not in so desperate a situation, as we have seen 
that there are genuine reasons for us to believe that our 
wait for the New Physics is about to end.    The next five years or so hold rich
promise for a significant improvement of our knowledge of $\gamma/\phi_3$ and $\theta_{13}$,
and our prospects of uncovering something unforeseen through the study of $B^0 \to K^{*0}\ell^+\ell^-$ and,
perhaps, in charm decays.  Most excitingly, the opening act of the LHC era
(in parallel with the closing scenes of the Tevatron programme) have the very real
possibility of revealing non-Standard Model contributions in both the measurement
of $\phi_s$ and the search for $B^0_s \to \mu^+\mu^-$.   When the New Physics is 
seen, the next challenge will be to establish its nature -- the question then
becomes `{\it who} is Godot?'

\section*{Acknowledgements}

I wish to thank the organisers for arranging a very stimulating and well run 
conference in a beautiful location and my fellow speakers for providing 
excellent and thought provoking talks.

\end{document}